\RequirePackage{rotating}
\documentclass[a4paper,fleqn,usenatbib]{mnras}

\usepackage[T1]{fontenc}
\usepackage{ae,aecompl}

\usepackage{graphicx}	
\usepackage{rotating}  
\usepackage{amsmath}	
\usepackage{amssymb}	
\usepackage{pifont}
\usepackage{caption} 

\newcommand{\com}[1]{\textnormal{#1}}

\title[Breaking formalism intrinsic degeneracies]{Generalised model-independent characterisation of strong gravitational lenses VI: the origin of the formalism intrinsic degeneracies and their influence on $H_0$}

\author[J. Wagner]{
Jenny Wagner$^{1}$\thanks{E-mail: j.wagner@uni-heidelberg.de}
\\
$^{1}$Universit\"at Heidelberg, Zentrum f\"ur Astronomie, Astronomisches Rechen-Institut, M\"onchhofstr. 12--14, 69120 Heidelberg, Germany\\
}

\date{Accepted XXX. Received YYY; in original form ZZZ}

\pubyear{2019}

\begin{document}
\label{firstpage}
\pagerange{\pageref{firstpage}--\pageref{lastpage}}
\maketitle

\begin{abstract}
We give a physical interpretation of the formalism intrinsic degeneracies of the gravitational lensing formalism that we derived on a mathematical basis in part IV of this series. 
We find that all degeneracies occur due to the partition of the mass density along the line of sight. 
Usually, it is partitioned into a background (cosmic) density and a foreground deflecting object. 
The latter can be further partitioned into a main deflecting object and perturbers. 
Weak deflecting objects along the line of sight are also added, either to the deflecting object or as a correction of the angular diameter distances, perturbing the cosmological background density. 
A priori, this is an arbitrary choice of reference frame and partition.
They can be redefined without changing the lensing observables which are sensitive to the integrated deflecting mass density along the entire line of sight. 
Reformulating the time delay equation such that this interpretation of the degeneracies becomes easily visible, we note that the source can be eliminated from this formulation, which simplifies reconstructions of the deflecting mass distribution or the inference of the Hubble constant, $H_0$. 
Subsequently, we list necessary conditions to break the formalism intrinsic degeneracies and discuss ways to break them by model choices or including non-lensing observables like velocity dispersions along the line of sight with their advantages and disadvantages.
We conclude with a systematic summary of all formalism intrinsic degeneracies and possibilities to break them.
\end{abstract} 

\begin{keywords}
cosmology: distance scale -- gravitational lensing: strong -- gravitational lensing: weak -- methods: analytical
\end{keywords}



\section{Introduction}
\label{sec:introduction}

In the first three parts of this paper series, \cite{bib:Wagner1}, \cite{bib:Wagner2}, and \cite{bib:Wagner3}, we have developed a method to determine local lens properties from measured observables in multiple images without assuming a specific model for the gravitational lens, like a specific mass density distribution or deflection potential. 
In the fourth part, \cite{bib:Wagner4}, we derived the most general class of invariance transformations of the gravitational lensing formalism that leave the observable properties of multiple images unchanged from a purely mathematical point of view. These degeneracies occur for \emph{any} lens reconstruction approach based on the general gravitational lensing equations.
They are independent of the lens model specified, may it be a parametric mass density profile or a free-form ansatz consisting of basis functions. 
They are also independent of the statistical viewpoint how to define the optimisation function, may the frequentist's or Bayesian statistics be employed.
In \cite{bib:Jullo}, \cite{bib:Liesenborgs}, \cite{bib:Keeton}, \cite{bib:Saha}, \cite{bib:Grillo2}, \cite{bib:Zitrin}, or \cite{bib:Merten} some approaches and implementations of lens reconstructions can be found.
Each of them treats and breaks the degeneracies slightly differently, based on their assumptions, especially the regularisation constraints.

In the lensing formalism, as it is described e.g. in \cite{bib:SEF} or \cite{bib:Petters}, the time delay equation is of great interest to determine the Hubble constant, $H_0$, from observed differences in the arrival times of light from multiple images of a time-varying background source, see e.g. \cite{bib:Refsdal}, \cite{bib:Suyu}, \cite{bib:Grillo}. 
Therefore, knowing the degeneracies of this equation is highly important to determine the width of the confidence bounds on $H_0$. 
The latter enters into the time delay equation through the cosmological standard model, which sets a background on top of which gravitational lensing is modelled and provides model-based angular diameter distances between the observer, the lens, and the source. 
Thus, embedding gravitational lensing into a cosmological background, it cannot be described independently of this background. 
In turn, we want to use observations of multiple images caused by a strong gravitational lens to infer parameters of the cosmological background model, see e.g.\ \cite{bib:Collett},  \cite{bib:Rasanen}, \cite{bib:Suyu}, and \cite{bib:Magana}, for some methods.

To determine $H_0$, current algorithms employ lens modelling approaches as mentioned above to calculate the ratio of angular diameter distances occurring in the time delay equation. 
The angular diameter distances are defined in such a way that the time delay equation has the same form in various cosmological models. 
Hence, having determined the (ratio of) angular diameter distances, any cosmological parameter inference can be performed on the angular diameter distances. 

To investigate the influence of the cosmological background model on the time delay difference between multiple images, in \cite{bib:Wagner5}, we assumed a Friedmann-Lemaître-Robertson-Walker cosmological model (FLRW model) and set up a distance measure for the angular diameter distances that is based on standardisable supernovae. 
Consequently, the distances were independent of a specific parametrisation of the FLRW model. 
As also stated in \cite{bib:Scolnic}, supernova data sets only constrain the expansion function of the universe up to an overall distance scale, leaving $H_0$ as the only free parameter of the FLRW model\footnote{Alternatively, the absolute magnitude $M$ that standardises the supernovae could be used, as $H_0$ and $M$ are mutually dependent.}. 
We calculated the relative precision of the ratio of angular diameter distances that enters the time delay equation and concluded that a parametrised expansion function, as inferred from the CMB by \cite{bib:Planck} or as fitted to the Pantheon sample of supernovae by \cite{bib:Scolnic} agrees to a parameter-free, data-based expansion function within the confidence intervals.
Hence, without a significant loss in precision, our approach is now independent of any Friedmann parametrisation of the cosmological background model with $H_0$ as the only free parameter left.  

In this sixth part of the paper series, we combine the results of \cite{bib:Wagner3}, \cite{bib:Wagner4}, and \cite{bib:Wagner5} to give an encompassing physical explanation for the degeneracies occurring in the standard gravitational lensing formalism. We particularly focus on the impact of the degeneracies for the determination of $H_0$ from the time delay equation.
In addition, we investigate ways to break the degeneracies by adding further observables or using model assumptions. 
Supported by numerous works of other authors, e.g. \cite{bib:SEF}, \cite{bib:Liesenborgs1}, \cite{bib:Schneider2}, \cite{bib:Xu_H0}, \cite{bib:Sonnenfeld}, to name a few, we hope that this part of the paper series will 
not only settle the questions about potentially arising degeneracies in our approach  but also contribute to explain the degeneracies and biases that might be expected in other approaches.

The next sections are organised as follows: 
In Section~\ref{sec:principles}, we review the general principles behind the gravitational lensing degeneracies, as already noted in \cite{bib:SEF}. 
We recapitulate the basics of gravitational lensing again to highlight all approximations and assumptions that lead to the equations usually employed. Alongside, we review the current status of observational support for the assumptions and approximations.
Based on these prerequisites and definitions, we discuss the ambiguities in the time delay equation and their impact on the determination of $H_0$ in Section~\ref{sec:solutions}. We give necessary conditions to break the degeneracy between the Fermat potential and $H_0$ and investigate how lens models break it. 
As an alternative, we discuss the coupling of the time delay equation with the Jeans equation that relates observed velocity dispersions to derivatives of the local, three-dimensional gravitational potential. Our main findings concerning the advantages and disadvantages of both methods to break the degeneracy are summarised in Section~\ref{sec:summary}. 
Section~\ref{sec:conclusion} concludes with a diagrammatic overview of all degeneracies of gravitational lensing and possibilities to break them.

\section{Formalism intrinsic origins of degeneracies}
\label{sec:principles}

\subsection{Origin of degeneracies}
\label{sec:theory}

\begin{figure*}
\centering
\includegraphics[width=0.9\textwidth]{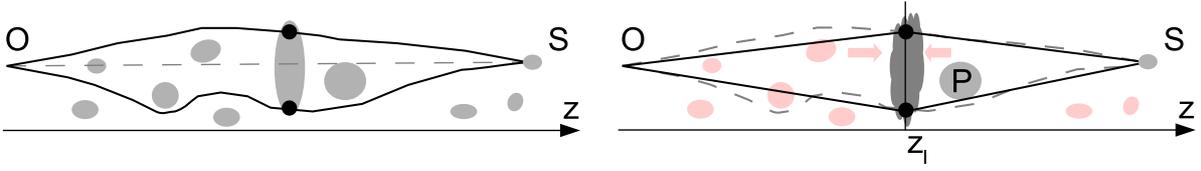}
\caption{Left: Propagation of light rays (marked by the black, solid lines) from a source $S$ to an observer $O$ through an inhomogeneous mass density distribution along the line of sight (marked by the grey objects) on top of a homogeneous and isotropic background. The unperturbed propagation through the background mass density is marked by the grey, dashed line, observations of multiple images are marked by the black dots; Right: Effective gravitational lensing description by projecting the small-scale mass densities along the line of sight (marked in red) onto the lens plane at $z_\mathrm{l}$, where the largest mass density is located. The mass density marked by "P" can be treated as a perturbation of the deflecting mass density at $z_\mathrm{l}$. The observations (marked by the black dots) are invariant to the description and therefore located at the same positions.}
\label{fig:mass_density_distribution}
\end{figure*}

Following the derivation of \cite{bib:SEF} and using the notation introduced in \cite{bib:Wagner5}, we start with the assumptions of an isotropic and spatially homogenous universe on large scales in a fundamental frame, i.e. for observers that are comoving to an overall, undetectable mean motion. This universe can be described by a Robertson-Walker metric with a line element given by
\begin{align}
\mathrm{d}s^2 &= - c^2 \mathrm{d}t^2 + a(t)^2 \mathrm{d} \boldsymbol{l}^2  \;.
\label{eq:RW_metric}
\end{align}
Light rays that propagate along null-geodesics, i.e. for which $\mathrm{d}s^2 = 0$, in this model are called unperturbed light rays. 
Adding an inhomogenously distributed mass density to the homogeneous and isotropic background mass density, the light rays become perturbed along their paths. 
The perturbation is described as follows: 
The light coming from the source propagates through the spatially homogeneous and isotropic universe and is deflected from its unperturbed path by the mass density distribution along its path.
Depending on the observed luminous matter density distribution along this path, the entire mass density distribution can be partitioned into the following components:  
(0) the homogeneous and isotropic background mass density, 
(1) a deflecting mass density in a locally confined region on top of the background (for instance a galaxy or a galaxy cluster), 
(2) potentially smaller, perturbing mass densities located in the vicinity of the main deflecting mass density (referred to as satellites in the following), 
(3) smaller inhomogeneities along the light path not belonging to the deflecting nor the satellite mass densities. 
Figure~\ref{fig:mass_density_distribution} (left) sketches the situation.

This partition of the mass density distribution along a path of a light ray is often too detailed, as we do not have observations at each point along the light path to characterise the mass density point-wise. 
Thus, we usually base our theoretical description on the assumption that we can divide the mass density into a homogeneous and isotropic background and one or several two-dimensional lens planes onto which we project the inhomogeneous mass density not belonging to the background. 
\cite{bib:McCully} gives an encompassing overview of the currently pursued partitioning approaches. 
Figure~\ref{fig:mass_density_distribution} (right) illustrates an effective theory for the example mass density distribution of Figure~\ref{fig:mass_density_distribution} (left), assuming we only had the observation of two multiple images at redshift $z_\mathrm{l}$.
As a result, the three-dimensional morphometry of the deflecting mass density is lost in the projection and we characterise the effectively deflecting mass density distribution(s) within the lens plane(s) by the observables. 
In \cite{bib:Wagner1} and \cite{bib:Wagner2}, we derived the effective local lens properties that can be determined in a single lens plane using one set of multiple images of a background source, as depicted in Figure~\ref{fig:mass_density_distribution} (right). 

Given the sparsity of multiple image observations, the effective description of the gravitational lensing configuration has to be chosen accordingly in order not to introduce variables that are not constrained by observations. 
But, even reducing the effective description to the minimum amount of degrees of freedom, several partitions are still possible that all give rise to formalism intrinsic degeneracies:
(a) In an isotropic and homogeneous universe with a single deflecting mass density, we are free to set the background mass density to an arbitrary value and subsequently adapt the deflecting mass density, such that the overall mass density that causes the multiple images remains the same.
(b) In an isotropic, homogeneous universe with a deflecting mass density and small-scale inhomogeneities distributed along the line of sight, we are free to absorb these inhomogeneities into the background by perturbing the background metric. This redefines the (angular diameter) distances. Alternatively, we can project the small-scale inhomogeneities onto the lens plane(s).
(c) Adding a second mass density into (b) and assuming that it is located in the proximity of the first deflecting mass density, we are free to redistribute any mass density between the background, the main lens, and the satellite as long as the observed multiple images remain invariant.

In the following, we will rederive the equations of the standard gravitational lensing formalism and relate the degeneracies of (a)--(c) to transformations of the respective variables. 
Since we already treated the case (c) of a main lens with a satellite in \cite{bib:Wagner3}, we focus on an effective description with a single lens plane as the minimal example to demonstrate the degeneracies.

\subsection{Degeneracies and invariance transformations}
\label{sec:derivation}

\subsubsection{Distances and $H_0$ in the cosmic background}

Starting from the background metric given by Equation~\eqref{eq:RW_metric}, distances from the observer to the lens plane at $z_\mathrm{l}$ (or $a_\mathrm{l}$, respectively\footnote{Cosmological considerations usually employ the scale factor of the metric, $a(t)$, while gravitational lensing uses the redshift $z$. In the following, we will employ both variables, which are related by $1+z = 1/a$, setting a=1 and z=0 today at $t_0$.}), where we assume an observation of a set of multiple images, are determined by $\mathrm{d}s^2 = 0$. Thus, we have to solve
\begin{align}
\mathrm{d} \boldsymbol{l} = - \dfrac{c \mathrm{d}t}{a(t)} \;,
\label{eq:proper_distance_differential}
\end{align}
in which the minus on the right-hand side is introduced to obtain larger distances for longer light travel times. 
In the following, we denote the three-dimensional position in space as $\boldsymbol{l}$, while $l$ is the absolute value of that distance.
We assume that $a(t)$ fulfils the Friedmann equations
\begin{align}
H(a)^2 \equiv \left( \dfrac{\dot{a}}{a}\right)^2 = \dfrac{8\pi G}{3} \rho(a) - \dfrac{K c^2}{a^2} + \dfrac{\Lambda c^2}{3} \;,
\label{eq:Friedmann_equation}
\end{align}
in which $\rho(a)$ denotes the matter and radiation content of the universe, $K$ the spatial curvature, and $\Lambda$ the cosmological constant. 
As usual, $G$ is the gravitational constant and $c$ the speed of light. 
Separating $H(a)$ into a normalisation constant, $H_0$, and the cosmic expansion function, $E(a)$, as also done in \cite{bib:Wagner4},
\begin{align}
H(a)^2 \equiv H_0^2 E(a)^2 = H_0^2 \left( \dfrac{8\pi G}{3} \dfrac{\rho(a)}{H_0^2} - \dfrac{K c^2}{H_0^2 a^2} + \dfrac{\Lambda c^2}{3 H_0^2} \right) \;,
\label{eq:Ha}
\end{align}
the normalisation condition $E(1) = 1$ relates $H_0$ to the total energy content of the universe today, at $a(t_0)=1$, by 
\begin{align}
H_0 = \sqrt{ \dfrac{8\pi G}{3} \rho(1) - K c^2+ \dfrac{\Lambda c^2}{3}} \;,
\label{eq:H0_density}
\end{align}
so that we can interpret a rescaling of $H_0$ as a rescaling of the energy content of the universe today.
Inserting Equation~\eqref{eq:Ha} into Equation~\eqref{eq:proper_distance_differential}, we arrive at the proper distance to the lens plane at $a_\mathrm{l}$
\begin{align}
d_{OL} = \dfrac{c}{H_0} \int \limits_{a_\mathrm{l}}^{1} \dfrac{\mathrm{d}a}{a E(a)} \equiv \dfrac{\tilde{d}_{OL}}{H_0}  \;.
\label{eq:D_prop} 
\end{align}

In order to improve distance estimates from measured redshifts, approaches have been developed to refine the assumption of a spatially homogeneous matter density distribution, among others, see \cite{bib:Dyer}, \cite{bib:Dyer1}, \cite{bib:Bonvin}, \cite{bib:Bolejko2}, or \cite{bib:Bolejko4}. 
Most importantly, for the degeneracies arising in the gravitational lensing formalism, \cite{bib:Bolejko} showed that the distance-redshift relation as defined by Equation~\eqref{eq:D_prop} holds for spatially homogeneous universes and is a good approximation with negligible corrections for universes in which density perturbations along a line of sight average out. 
If a line of sight has a non-zero mean of density perturbations, which can occur in case (b) of Section~\ref{sec:theory}, then, a modified Dyer-Roeder approximation or the linear approximation approach of weak lensing can be employed equally well to describe the perturbed light paths. 
Alternatively, as discussed in \cite{bib:Keeton1}, \cite{bib:McCully}, \cite{bib:Birrer4}, and \cite{bib:Wong1}, the line-of-sight structures can be effectively accounted for as a sheet of an external mass density, either in a multi-lens-plane approach or projected into a single lens plane.

Simulations that investigate the abundance and distribution of line-of-sight inhomogeneities are detailed in \cite{bib:Despali}. 
Furthermore, relative abundances and degeneracies with respect to satellites are also discussed in this work.
Observational investigations about structures along the line of sight are detailed in \cite{bib:Wong} and references therein. 
\com{Including this information in the lens modelling increases the precision of the resulting lens properties. At the same time, the increasing amount of degrees of freedom in the description has to be constrained and the additional degeneracies must be broken or carefully analysed.}

Hence, the derivations of this section yield that $H_0$ is strongly related to the entire energy content of the universe today by Equation~\eqref{eq:H0_density} and that it is possible to perturb either the background metric, including small-scale inhomgeneities, or to treat them as weakly deflecting mass densities. 
Clearly, fixing $H_0$ with a different normalisation condition than Equation~\eqref{eq:H0_density}, $E(a)$ can be rescaled in the same way, such that the distance measures like Equation~\eqref{eq:D_prop} remain invariant. 
Assuming a Dyer-Roeder extension of the background to an on-average spatially homogeneous universe with a smoothness parameter, by which the distances are scaled due to the inhomogeneities, we note that the different normalisation of $H_0$ could be cancelled by choosing the smoothness parameter accordingly. 
As was demonstrated in \cite{bib:Bolejko}, a redshift-independent, global smoothness parameter is not sufficient to include the inhomogeneities in the distance-redshift relation. 
But it is possible to find a redshift-dependent parameter to bring the weak lensing and the Dyer-Roeder approximation into agreement.
Setting up this equivalence of describing the background is a topic on its own. 
As all approaches to gravitational lensing assume a FLRW metric, we do not consider this degeneracy within the background description further. 

\begin{figure}
\centering
\includegraphics[width=0.33\textwidth]{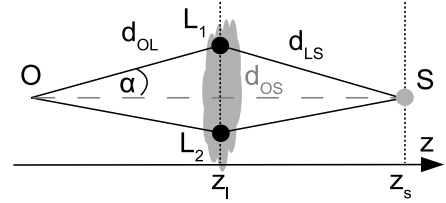}
\caption{Propagation of an unperturbed light ray from the source $S$ at redshift $z_\mathrm{s}$ to the observer $O$ along the grey, dashed line and propagation of two perturbed light rays from $S$ over $L_1$ or $L_2$ to $O$ along the black, solid paths. The perturbation of the background mass density is assumed to be effectively described as a projected, two-dimensional mass density at redshift $z_\mathrm{l}$, marked in grey behind $L_1$ and $L_2$.}
\label{fig:deflection}
\end{figure}

\newpage
\subsubsection{Time delay differences due to inhomogeneities}
\label{sec:inhomogeneities}

Next, we consider a single projected, two-dimensional mass density distribution in a lens plane at $z_\mathrm{l}$ on top of a spatially homogeneous and isotropic background given by Equation~\eqref{eq:RW_metric}. 
We assume that this mass density generates multiple images of a background source located at $z_\mathrm{s}$. 
Furthermore, we assume that the lens, the source, and the observer do not move relative to each other and that space is flat, i.e. $K=0$. For $K \ne 0$, the calculations are completely analogous and lead to the same result due to the definition of distances.
Figure~\ref{fig:deflection} illustrates the situation for an example configuration of two multiple images.
Without loss of generality, the statements are valid for any number of multiple images from the same source.
Following \cite{bib:Alchera1}, the time delay $t$ for a light ray on a perturbed path arriving at the observer at time $t_\mathrm{p}$ compared to the unperturbed path arriving at the observer at time $t_\mathrm{u}$ is given by
\begin{align}
t \equiv \int \limits_{t_\mathrm{u}}^{t_\mathrm{p}} \dfrac{\mathrm{d}t}{a(t)}= \dfrac{\left| \boldsymbol{d}_{OL} \right| + \left| \boldsymbol{d}_{LS} \right| - \left| \boldsymbol{d}_{OS} \right| }{c} \;,
\label{eq:time_delay}
\end{align}
using the notation of Figure~\ref{fig:deflection}. 
Assuming that the time delay is much smaller than the Hubble time $t \ll H_0^{-1}$, we can approximate the integral on the left-hand side by 
\begin{align}
t \approx \dfrac{t_\mathrm{p}-t_\mathrm{u}}{a(t_0)} = t_\mathrm{p}-t_\mathrm{u} \;.
\label{eq:time_delay2}
\end{align}
The distances on the right-hand side are the proper lengths from one point $i$ in space to another point $j$ as indicated in Figure~\ref{fig:deflection} for the distances between the observer, the lens, and the source. 
Then, the distances along the line of sight starting at $O$ are given by
\begin{align}
d_{ij} = \int \limits_i^j \mathrm{d}r \;, \quad i = O\;, \quad j=L, S\;,
\label{eq:distances}
\end{align}
in which $r$ denotes the radial coordinate with origin $O$. 
$d_{LS}$ is given by
\begin{align}
d_{LS} = \sqrt{\boldsymbol{d}_{OL}^2 + \boldsymbol{d}_{OS}^2 - 2 \boldsymbol{d}_{OL} \boldsymbol{d}_{OS}} \;.
\label{eq:dms}
\end{align}
Assuming small angles $\boldsymbol{\alpha}$ between the vectors $\boldsymbol{d}_{OL} $ and $\boldsymbol{d}_{OS}$\footnote{We introduce $\boldsymbol{\alpha}$ as a vector, as it spans the angle between $\boldsymbol{d}_{OL} $ and $\boldsymbol{d}_{OS}$ in the plane orthogonal to the line of sight. This will become relevant in the next sections.}, we find
\begin{align}
t \approx \dfrac{d_{OS} d_{OL}}{c(d_{OS}-d_{OL})} \, \dfrac{\boldsymbol{\alpha}^2}{2} \approx \dfrac{d_{OS} d_{OL}}{c d_{LS}} \, \dfrac{\boldsymbol{\alpha}^2}{2} \;,
\label{eq:small_angles}
\end{align}
for $d_{OS} - d_{OL} \approx d_{LS}$.
Scaling the proper distances by $1/(1+z_i)$, $i=L, S$, we obtain the time delay between the perturbed and the unperturbed light ray in terms of the angular diameter distances as
\begin{align}
t \approx \dfrac{(1+z_\mathrm{l}) }{c}\dfrac{D_A(0,z_\mathrm{s}) D_A(0,z_\mathrm{l})}{D_A(z_\mathrm{l}, z_\mathrm{s})} \, \dfrac{\boldsymbol{\alpha}^2}{2} \;,
\end{align}
in which $D_A(z_i, z_j)$ denotes the angular diameter distance between the redshifts $z_i$ and $z_j$.
In \cite{bib:Alchera1}, the validity of the small angle approximation which leads to Equation~\eqref{eq:small_angles} was investigated for the CASTLES catalogue of lenses. It was found that including higher order contributions resulted in negligible contributions. 

Subtracting the time delay of the multiple image passing $L_2$ from the one passing $L_1$, we obtain the geometrical time delay difference between the multiple images
\begin{align}
\tau_\mathrm{g} = \dfrac{(1+z_\mathrm{l}) }{c}\dfrac{D_A(0,z_\mathrm{s}) D_A(0,z_\mathrm{l})}{D_A(z_\mathrm{l}, z_\mathrm{s})} \, \left( \dfrac{\boldsymbol{\alpha}_1^2}{2} -\dfrac{\boldsymbol{\alpha}_2^2}{2} \right) \;. 
\label{eq:geometric_tau}
\end{align}
This result was derived in \cite{bib:Gorenstein}. It was further discussed in \cite{bib:Wagner4}, in which we had defined
\begin{align}
\Delta \mathcal{G} =  \dfrac{\boldsymbol{\alpha}_1^2}{2} -\dfrac{\boldsymbol{\alpha}_2^2}{2} \;.
\end{align}

From Equation~\eqref{eq:geometric_tau}, we read off the degeneracy stated in (b) in Section~\ref{sec:theory}. 
The angular diameter distances are measured along \emph{specific} lines of sight and the $\boldsymbol{\alpha}$ are the \emph{local} deflection angles of the light rays caused by the local deflecting mass density. 
Therefore, without knowing the mass density distribution along the line of sight, it is not possible to disentangle inhomogeneities included in the distance measure of a perturbed background metric from a contribution of a local deflecting mass density to $\boldsymbol{\alpha}$ projected into the lens plane.  
This degeneracy is merely a choice of background metric.
We choose to determine distances based on the FLRW metric and assume line-of-sight structures to be accounted for in the projected, two-dimensional mass density in the following. 
As stated in \cite{bib:Scolnic}, magnifications of the supernovae in the Pantheon sample due to lensing structures along the line of sight are considered as flux biases and calibrated out. 
Hence, the distance measures developed in \cite{bib:Wagner5} are based on a non-parametric, data-based FLRW metric, which is consistent with our choice to project deflecting masses along the line of sight into the lens plane.

The geometric part of the time delay difference is only one part of the observed time delay difference between two multiple images.
The time delay caused by the deflecting mass density at the lens plane also has to be taken into account. This part is called the Shapiro delay. 
To determine the Shapiro delay, we linearly perturb the FLRW metric by a deflection potential $\phi(t,\boldsymbol{l})$, such that
\begin{align}
\mathrm{d}s^2 = - \left(1 + \dfrac{2\phi(t,\boldsymbol{l})}{c^2} \right) \, c^2 \mathrm{d}t^2 + a(t)^2 \left(1-\dfrac{2\phi(t,\boldsymbol{l})}{c^2}\right) \, \mathrm{d} \boldsymbol{l}^2 \;.
\end{align}
Then, the difference in the travel times between an unperturbed and a perturbed light ray is
\begin{align}
t = \int \limits_{t_\mathrm{u}}^{t_\mathrm{p}} \dfrac{\mathrm{d}t}{a(t)} \approx \dfrac{t_\mathrm{p}-t_\mathrm{u}}{a(t_0)} = t_\mathrm{p}-t_\mathrm{u} \;,
\end{align}
in which we assume in the last steps that the delay is smaller than the Hubble time, as in Equation~\eqref{eq:time_delay2}.
Analogously to Equation~\eqref{eq:time_delay}, we obtain from $\mathrm{d}s^2 = 0$ 
\begin{align}
t = \dfrac{1}{c} \int \limits_{\gamma_{L_1}} \mathrm{d} \boldsymbol{l} \sqrt{\dfrac{1-\tfrac{2\phi(t,\boldsymbol{l})}{c^2}}{1+\tfrac{2\phi(t,\boldsymbol{l})}{c^2}}} \approx  \dfrac{1}{c} \int \limits_{\gamma_{L_1}} \mathrm{d} \boldsymbol{l}  \left(1 - \dfrac{2\phi(t,\boldsymbol{l})}{c^2} \right) \;.
\end{align}
For the last step, we use that the perturbation is assumed to be small, such that $2\phi(t,\boldsymbol{l})/c^2 \ll 1$. $\gamma_{L_1}$ denotes the path of a light ray passing the multiple image $L_1$ in Figure~\ref{fig:deflection}.
Thus, the time delay difference between the two light rays passing $L_1$ and $L_2$ in Figure~\ref{fig:deflection} caused by a perturbing potential is given by
\begin{align}
\tau_\mathrm{s} = - \dfrac{2}{c^3} \left( \int \limits_{\gamma_{L_1}} \mathrm{d} \boldsymbol{l} \, \phi(t, \boldsymbol{l}) - \int \limits_{\gamma_{L_2}} \mathrm{d} \boldsymbol{l} \, \phi(t,\boldsymbol{l}) \right) \;.
\label{eq:shapiro_tau_general}
\end{align}
Inserting our two-dimensional, projected deflection potential in the lens plane $\tilde{\psi}(\boldsymbol{\xi})$ with $\boldsymbol{\xi} \in \mathbb{R}^2$, i.e. $\phi(t,\boldsymbol{l}) = \tilde{\psi}(\boldsymbol{\xi})/a_\mathrm{l}$, the integration along the light path is reduced to the position of the lens plane along the line of sight. Thus, the Shapiro delay between the two light rays due to different potentials at $L_1$, located at $\boldsymbol{\xi}_1$, and $L_2$, located at $\boldsymbol{\xi}_2$ in the lens plane, reads
\begin{align}
\tau_\mathrm{s} = - \dfrac{2}{a_\mathrm{l} \, c^3} \left( \tilde{\psi}(\boldsymbol{\xi}_1) - \tilde{\psi}(\boldsymbol{\xi}_2) \right) \;, \quad \boldsymbol{\xi}_{1}, \boldsymbol{\xi}_{2} \in \mathbb{R}^2 \;.
\label{eq:shapiro_tau}
\end{align}
Adding Equations~\eqref{eq:geometric_tau} and \eqref{eq:shapiro_tau}, we obtain the total time delay difference between two light rays passing $L_1$ and $L_2$. 

Hence, as already stated in Section~\ref{sec:theory}, the only degeneracy arising in this part of the time delay difference occurs in the definition of the projected potential $\tilde{\psi}(\boldsymbol{\xi})$. 
Assuming an FLRW metric and projecting all inhomogeneities along the line of sight onto the lens plane, different mass density configurations can lead to the same projected mass density, and thus to the same observed time delay difference between light rays coming from multiple images. 
Extending the approach from a lens plane to a volume of deflection, as done in \cite{bib:Alchera1}, requires additional assumptions and observables to track the light path through this volume.
Such extensive observations have, for instance, been carried out for the galaxy-scale lens HE 0435-1223, as detailed in \cite{bib:Sluse}, \cite{bib:Rusu}, \cite{bib:Tihhonova}.

\subsubsection{Scaling to dimensionless quantities}
\label{sec:scaling}

Since measurements are comparisons of observables to reference values, the quantities occurring in the lensing formalism are usually scaled by reference values. In the gravitational lensing formalism, the dimensionless quantities are the observed angular positions of the multiple images and derived angular quantities like the angular position of the source or the deflection angle caused by the mass density in the lens plane, see e.g. \cite{bib:SEF}. 

We start by scaling the two-dimensional distances $\boldsymbol{\xi}$ from the origin of the coordinate system in the lens plane to the multiple images.
Subsequently, we scale the source position $\boldsymbol{\eta}$, such that we obtain the following angular positions on the celestial sphere in the lens and source plane
\begin{align}
\boldsymbol{x} = \dfrac{\boldsymbol{\xi}}{D_\mathrm{l}} \;, \quad \boldsymbol{y} = \dfrac{\boldsymbol{\eta}}{D_\mathrm{s}} \;,
\label{eq:scaling}
\end{align}
using the notation of \cite{bib:SEF}.
Then, the deflection angle $\boldsymbol{\alpha}(\boldsymbol{x})$ is determined by 
\begin{align}
\boldsymbol{\alpha}(\boldsymbol{x}) =\boldsymbol{x} - \boldsymbol{y} \;.
\label{eq:alpha}
\end{align}

Next, we scale the two-dimensional deflection potential introduced in Section~\ref{sec:inhomogeneities}. To do so, we add Equations~\eqref{eq:geometric_tau} and \eqref{eq:shapiro_tau} to obtain
\begin{align}
\tau &= \tau_\mathrm{g} + \tau_\mathrm{s}  \\
 &= \dfrac{1}{a_\mathrm{l} \, c}\dfrac{D_\mathrm{s} D_\mathrm{l}}{D_\mathrm{ls}} \, \left( \dfrac{\boldsymbol{\alpha}_1^2}{2} -\dfrac{\boldsymbol{\alpha}_2^2}{2} \right) -  \dfrac{2}{a_\mathrm{l} \, c^3} \left( \tilde{\psi}(\boldsymbol{\xi}_1) - \tilde{\psi}(\boldsymbol{\xi}_2) \right) \\
 &=  \dfrac{1}{a_\mathrm{l} \, c}\dfrac{D_\mathrm{s} D_\mathrm{l}}{D_\mathrm{ls}} \, \left( \dfrac{\boldsymbol{\alpha}_1^2}{2} -\dfrac{\boldsymbol{\alpha}_2^2}{2} - \dfrac{2}{c^2} \dfrac{D_\mathrm{ls}}{D_\mathrm{l} D_\mathrm{s}} \left( \tilde{\psi}(\boldsymbol{\xi}_1) - \tilde{\psi}(\boldsymbol{\xi}_2) \right) \right) \;,
\end{align}
in which we abbreviated the angular diameter distances $D_A(z_i,z_j)$ by $D_{ij}$. 
Defining the scaled deflection potential
\begin{align}
\psi(\boldsymbol{\xi}) \equiv \dfrac{2}{c^2} \dfrac{D_\mathrm{ls}}{D_\mathrm{l} D_\mathrm{s}} \tilde{\psi}(\boldsymbol{\xi}) \;,
\label{eq:scaled_potential}
\end{align}
and rescaling the lengths $\boldsymbol{\xi}$ using Equation~\eqref{eq:scaling}, we obtain
\begin{align}
\tau =  \dfrac{1}{a_\mathrm{l} \, c}\dfrac{D_\mathrm{s} D_\mathrm{l}}{D_\mathrm{ls}} \, \left[ \dfrac12 \left( \boldsymbol{\alpha}(\boldsymbol{x}_1)^2- \boldsymbol{\alpha}(\boldsymbol{x}_2)^2 \right) - \psi(\boldsymbol{x}_1) + \psi(\boldsymbol{x}_2) \right] \;.
\label{eq:time_delay3}
\end{align}
The first factors contain the geometry of the configuration, while the factor in squared brackets contains angular (scaled) quantities that are agnostic about the geometry. 
Defining the deflection potential as done in Equation~\eqref{eq:scaled_potential} automatically ensures that $\boldsymbol{\alpha}(\boldsymbol{x})$ in Equation~\eqref{eq:alpha} is the gradient of the deflection potential with respect to $\boldsymbol{x}$: $\boldsymbol{\alpha}(\boldsymbol{x}) = \nabla_{\boldsymbol{x}} \psi(\boldsymbol{x})$. 
Deviations from $\nabla \times \boldsymbol{\alpha}(\boldsymbol{x}) = 0$ could potentially arise, which amount to a degeneracy if they are smaller than the measurement precision. 
This kind of degeneracy in the deflection angles was first described as a source position transformation (SPT) in \cite{bib:Schneider1}. 
In \cite{bib:Wagner4}, we re-interpreted the global SPT in the single-lens-plane formalism as a local effect of an additional lens plane. 
Keeping the time delay difference constant because its value is observed, we arrived at the conclusion that our model-independent approach is not subject to this degeneracy by construction.  
So far, the astrophysical effects that require an extension of the lensing formalism to use deflection angles with $\nabla \times \boldsymbol{\alpha}(\boldsymbol{x}) \ne 0$ are rarely observed.
Therefore, we the focus on exact degeneracies here and only consider deflection angles derived from a deflection potential. 

Linking the deflection potential to the two-dimensional mass density distribution $\Sigma(\boldsymbol{x})$, we can also set up a scaled two-dimensional ``angular" mass density distribution, usually called convergence $\kappa(\boldsymbol{x})$. 
To do so, we require that $\psi(\boldsymbol{x})$ fulfils a Poisson equation\footnote{The mathematically thorough prerequisites for existence and uniqueness of solutions are detailed in \cite{bib:Wagner4}. 
To solve this two-dimensional Poisson equation, it is necessary to introduce the dimensionless variable $\boldsymbol{x}$, as the Green's function of the two-dimensional Laplace operator is logarithmic in $\boldsymbol{x}$.} as
\begin{align}
\Delta_{\boldsymbol{x}} \psi(\boldsymbol{x}) = 2 \kappa(\boldsymbol{x}) \;.
\label{eq:Poisson_x}
\end{align} 
By inserting $\psi(\boldsymbol{\xi})$ of Equation~\eqref{eq:scaled_potential} and obeying the transformation of Equation~\eqref{eq:scaling} when determining the Laplace operator with respect to $\boldsymbol{x}$, we obtain
\begin{align}
\Delta_{\boldsymbol{x}} \psi(\boldsymbol{x}) = D_\mathrm{l}^2 \dfrac{2}{c^2} \dfrac{D_\mathrm{ls}}{D_\mathrm{l} D_\mathrm{s}} \Delta_{\boldsymbol{\xi}} \tilde{\psi}(\boldsymbol{\xi}) \;.
\label{eq:Poisson_psi}
\end{align}
As $\tilde{\psi}(\boldsymbol{\xi})$ is a weak, linear perturbation to the FLRW metric (see Section~\ref{sec:inhomogeneities}), it fulfils the Newtonian limit of Einstein's field equations. Hence, it fulfils the three-dimensional Poisson equation, with $\boldsymbol{l} \in \mathbb{R}^3$,
\begin{align}
\Delta_{\boldsymbol{l}} \left( \tilde{\psi}(\boldsymbol{\xi}) \, \delta(z-z_\mathrm{l}) \right) = 4 \pi G \Sigma(\boldsymbol{\xi}) \, \delta(z-z_\mathrm{l})\;, \quad 
\label{eq:Newtonian_limit}
\end{align}
which is only non-zero in the lens plane.
Inserting Equation~\eqref{eq:Newtonian_limit} into Equation~\eqref{eq:Poisson_psi}, we replace $\Sigma(\boldsymbol{\xi})$ by $\Sigma(\boldsymbol{x})$ to obtain
\begin{align}
\Delta_{\boldsymbol{x}} \psi(\boldsymbol{x}) = D_\mathrm{l}^2 \dfrac{2}{c^2} \dfrac{D_\mathrm{ls}}{D_\mathrm{l} D_\mathrm{s}} 4 \pi G \Sigma(\boldsymbol{x}) \;.
\label{eq:Poisson_equation}
\end{align}
The definition for $\kappa(\boldsymbol{x})$ is obtained by comparing the result to Equation~\eqref{eq:Poisson_x}
\begin{align}
\kappa(\boldsymbol{x}) \equiv \dfrac{\Sigma(\boldsymbol{x})}{\Sigma_0} \;, \quad \Sigma_0 \equiv \dfrac{c^2}{4 \pi G} \dfrac{D_\mathrm{s}}{D_\mathrm{l} D_\mathrm{ls}} \;.
\end{align}
We note that the background mass density is comprised in $\Sigma_0$ because $\Sigma_0$ is proportional to $H_0$ and contains the distance-redshift relations of the chosen background metric.

Thus, scaling to dimensionless quantities introduces a geometric degeneracy not yet mentioned. 
Considering the lensing equation, Equation~\eqref{eq:alpha}, we notice that it relates only angular positions to each other. Consequently, without fixing the distance-redshift relation, the lens and source planes can be relocated along the line of sight, as long as all relative observed angular image positions remain invariant
\begin{align}
\boldsymbol{x}_1 - \boldsymbol{x}_2 = \boldsymbol{\alpha}(\boldsymbol{x}_1) - \boldsymbol{\alpha}(\boldsymbol{x}_2) \;,
\label{eq:observables}
\end{align}
as also stated in \cite{bib:Schneider1}.

In \cite{bib:Wagner2}, \cite{bib:Wagner6}, we only employed observables of multiple images in the lens plane, like the positions of the centres of light or reference points within the images, or the quadrupoles of the intensity profiles around the centres of light. 
As a result, we obtained a lens characterisation in terms of reduced quantities, i.e. the reduced shear and ratios of convergences, or ratios of derivatives of the deflection potential at the critical curves. 
Putting these mathematically derived characteristics in the physical context of this section, we see that they are the maximum information retrievable from the scaled (angular) observables, as already noted in \cite{bib:Schneider5} and extended to higher orders by \cite{bib:Schneider6}.

\subsection{Summary of all degeneracies arising}
\label{sec:summary_of_degeneracies}

Summarising the results of Section~\ref{sec:derivation}, we find that the formalism intrinsic degeneracies of gravitational lensing originate due to two causes. 
The first one is the freedom to define a background cosmology with a redshift-distance relation that underlies the gravitationally lensing mass density distributions.
The second one is the geometric freedom to relocate the lens and source planes such that the angular positions observed in a multiple-image configuration and the differences in the arrival times of light measured between pairs of multiple images remain invariant. 
In \cite{bib:Wagner4}, we investigated the degeneracies arising in Equations~\eqref{eq:alpha} and \eqref{eq:time_delay3} for a given FLRW background cosmology with its standard distance-redshift relation.
In the following section, we discuss ways to break the degeneracies including further observations, with additional assumptions, and with inserting specific models for the deflecting mass density distribution. 

\section{Breaking the degeneracies}
\label{sec:solutions}

\subsection{Prerequisites}

\subsubsection{Observables from multiple images}
\label{sec:observables_mis}

We assume that the following observables of multiple images from an extended background source (like a galaxy) with a time-varying component (like a supernova, a fast radio burst, or a quasar with time-varying intensity) can be obtained:
the relative image positions between pairs of multiple images $\boldsymbol{x}_1 - \boldsymbol{x}_2$, measured between the centres of light of the extended images,
the quadrupole moments around  the centres of light of the extended images, or, alternatively, at least three reference points that can  be identified in all multiple images,  and the difference of the arrival times of pairs of multiple images $\tau$.

\subsubsection{Boundary conditions for the Poisson equation}
\label{sec:boundary_conditions}

Equation~\eqref{eq:Poisson_equation} relates the scaled mass density distribution $\kappa(\boldsymbol{x})$ at an angular position $\boldsymbol{x}$ to the deflection potential $\psi(\boldsymbol{x})$ at the same position. 
It requires boundary conditions to uniquely fix $\psi(\boldsymbol{x})$, as detailed in \cite{bib:Wagner4}.
As in \cite{bib:Wagner4}, we denote the lensing region by $\mathcal{X}$ and its boundary by $\partial \mathcal{X}$.
These boundary conditions introduce non-local contributions to the deflection angle or the deflection potential which can be physically interpreted as the deflection caused by parts of the mass density surrounding $\boldsymbol{x}$.
Equivalently, these non-local contributions can be accounted for by external shear or external convergence terms. 
Depending on the specific lensing problem, a characterisation based on the convergence, external convergence, and external shear may be suitable. 
Yet, we consider it easier and more efficient to employ the deflection potential $\psi(\boldsymbol{x})$ that comprises all these physical effects in a single scalar function.

Usually, the entire lens plane $\mathbb{R}^2$ is considered as $\mathcal{X}$, employing the gauge
\begin{align}
\left. \psi(\boldsymbol{x})\right|_{\partial \mathcal{X}} = 0 \;,
\label{eq:gauge}
\end{align}
i.e.\@ that the deflection potential  vanishes for $|\boldsymbol{x}| \rightarrow \infty$.
Yet, physically reasonable deflecting mass distributions are expected to be constrained to finite regions. \cite{bib:Saha1} already noted that adding a mass disk of finite extent, covering the area of all multiple images has the same effect on the observables as adding a mass sheet of infinite extent as suggested by \cite{bib:Falco}. 
For mass density distributions of finite extent fulfilling Equation~\eqref{eq:small_angles}, we can even impose Dirichlet boundary conditions on a finite domain $\mathcal{X}$ with a continuous, non-vanishing function $g$ on $\partial \mathcal{X}$
\begin{align}
\left. \psi(\boldsymbol{x})\right|_{\partial \mathcal{X}} = g(\boldsymbol{y}) \;, \quad  \boldsymbol{y}  = \left. \boldsymbol{x}\right|_{\partial  \mathcal{X}}\;,
\label{eq:non_vanishing_boundary}
\end{align}
without encountering additional degeneracies and without the need to consider the entire lens plane $\mathbb{R}^2$, as detailed in \cite{bib:Wagner4}. 
In the following, unless mentioned otherwise, we will employ Equation~\eqref{eq:gauge} but assume a general lensing domain $\mathcal{X}$ that need not be infinitely extended. 
The other case of a non-vanishing function as boundary condition, Equation~\eqref{eq:non_vanishing_boundary}, is left for future investigations when we couple the lensing information with other probes of the gravitational potential that provide such boundary conditions.

\subsection{Separating observables and unknowns in the time delay equation}
\label{sec:time_delay}

The right-hand side of the time delay equation, Equation~\eqref{eq:time_delay3}, does not contain any observed quantities, yet. Therefore, we reformulate it, as done in \cite{bib:Gorenstein} and \cite{bib:Wagner4} as
\begin{align}
\tau &=  \Gamma \, \left( \left(\boldsymbol{x}_1 - \boldsymbol{x}_2 \right)^\top  \dfrac{\boldsymbol{\alpha}(\boldsymbol{x}_1)+\boldsymbol{\alpha}(\boldsymbol{x}_2)}{2}  - \psi(\boldsymbol{x}_1) + \psi(\boldsymbol{x}_2) \right) \;, \label{eq:time_delay41}\\
&= \Gamma \Delta \phi(\boldsymbol{x}_1, \boldsymbol{x}_2) \;,
\label{eq:time_delay4}
\end{align}
with
$\Gamma = 1/(a_\mathrm{l} \, c) \, (D_\mathrm{s} D_\mathrm{l})/D_\mathrm{ls}$ and the differences between the Fermat potentials of the multiple images, $\Delta \phi(\boldsymbol{x}_1, \boldsymbol{x}_2)$.
Having fixed the background metric to be an FLRW metric, the angular diameter distances are defined accordingly, such that the degeneracy mentioned under (b) in Section~\ref{sec:theory} cannot occur anymore.
 Furthermore, given $H_0$, the degeneracy mentioned under (a) is also fixed, such that a measured time delay difference uniquely determines $\Delta \phi(\boldsymbol{x}_1, \boldsymbol{x}_2)$. 
Connecting the deflection potential $\psi(\boldsymbol{x})$ to a projected mass density $\kappa(\boldsymbol{x})$ and given some mathematical requirements on these functions, \cite{bib:Wagner4} showed that measuring the time delay for a given cosmological background model fixes $\kappa(\boldsymbol{x})$. 
The physical reason for this mathematically derived result can now be understood:
Fixing the background mass density distribution by selecting a metric and setting $H_0$ uniquely determines the two-dimensional deflecting mass density on top of it. 
Only the degeneracy between the terms in brackets remains, when $\Delta \phi(\boldsymbol{x}_1, \boldsymbol{x}_2)$ is divided into the geometric and the Shapiro delay. 
It implies that all deflection potentials $\psi(\boldsymbol{x})$ with $\boldsymbol{\alpha}(\boldsymbol{x}) = \nabla_{\boldsymbol{x}} \psi(\boldsymbol{x})$ fulfilling
\begin{align}
\psi (\boldsymbol{x}_1) - \psi (\boldsymbol{x}_2) - \left( \boldsymbol{x}_1 - \boldsymbol{x}_2 \right)^\top \dfrac{\boldsymbol{\alpha}(\boldsymbol{x}_1) + \boldsymbol{\alpha}(\boldsymbol{x}_2)}{2} = \dfrac{\tau}{\Gamma}
\label{eq:invariance_transformation}
\end{align}
are valid solutions.  
We note that the formulation of Equation~\eqref{eq:time_delay41} does not require to reconstruct the source, since the latter is completely defined by the multiple images and the deflecting mass density distribution. 
Writing the geometric part of the time delay as done in Equation~\eqref{eq:geometric_tau} greatly simplifies the study of the degeneracies.

Fixing the distance-redshift relation by a cosmological background model, usually an FLRW model, but leaving $H_0$ as a free parameter, all Fermat potentials and $H_0$ fulfilling
\begin{align}
\dfrac{ \Delta \phi(\boldsymbol{x}_1, \boldsymbol{x}_2)}{H_0} = \dfrac{\tau}{\tilde{\Gamma}}
\label{eq:H0_degeneracy}
\end{align}
are valid solutions. $\tilde{\Gamma}$ contains the scale-free angular diameter distances analogous to the right-hand side of Equation~\eqref{eq:D_prop}.
Inserting the distances based on the Pantheon sample into $\tilde{\Gamma}$, as set up in \cite{bib:Wagner5}, the right-hand-side of Equation~\eqref{eq:H0_degeneracy} is purely data-based.
As $H_0$ can be linked to a constant mass density by Equation~\eqref{eq:H0_density}, Equation~\eqref{eq:H0_degeneracy} proves that $ \Delta \phi(\boldsymbol{x}_1, \boldsymbol{x}_2)$ is only subject to a mass sheet degeneracy when $H_0$ is treated as a free parameter in the time delay equation. In other words, the degeneracy as defined in case (a) in Section~\ref{sec:theory} is not fixed in Equation~\eqref{eq:H0_degeneracy}.

\subsection{Necessary conditions to break the $H_0$-$\Delta \phi$-degeneracy}
\label{sec:breaking_degeneracies}

Equation~\eqref{eq:H0_degeneracy} contains the ratio of two unknowns, namely the difference between the Fermat  potentials at positions $\boldsymbol{x}_1$ and $\boldsymbol{x}_2$ and the Hubble constant $H_0$.
Hence, we require a second, independent equation to solve for $\Delta \phi(\boldsymbol{x}_1,\boldsymbol{x}_2)$ and $H_0$ separately.
This second equation can be 
(1) a direct measurement of $H_0$ or an assumption for it.
(2) a local observation or assumption constraining $\Delta \phi(\boldsymbol{x}_1,\boldsymbol{x}_2)$, which does not include $H_0$,
(3) a local observation or assumption constraining $\Delta \phi(\boldsymbol{x}_1,\boldsymbol{x}_2)$ and $H_0$ in a different way than Equation~\eqref{eq:H0_degeneracy}.
In the following, we will focus on (2) and (3) to analyse how the degeneracy in Equation~\eqref{eq:H0_degeneracy} can be broken to determine $H_0$.

Assuming that two time delay differences between a triple of multiple images coming from the same source are measured, the system of the two equations of Equation~\eqref{eq:H0_degeneracy} remains degenerate, as the deflection potentials for each image are scaled by the same factor to obey Equation~\eqref{eq:observables}.
Analogously, a $\chi^2$-parameter estimation of an ensemble of different multiple-image systems will yield a value for $H_0$ that is the weighted arithmetic mean of all $\Delta \phi \tilde{\Gamma}/\tau$ with the squared scaling factors of $\Delta \phi$ as weights. 
Thus, adding more time delay difference observations does not break  the degeneracy in Equation~\eqref{eq:H0_degeneracy} and the additional constraint between $\Delta \phi$ and $H_0$ in (3) must come from a different relation. 

\subsection{Breaking degeneracies by lens models}
\label{sec:model_breaking}

Inserting a lens model into Equation~\eqref{eq:invariance_transformation} means assuming a globally defined $\psi(\boldsymbol{x},\boldsymbol{p})$ depending on a vector of $n_\mathrm{p}$ parameters $\boldsymbol{p}$. Implicitly, a previously defined lens centre, e.g.\@ the symmetry centre of a galaxy or the brightest galaxy in a cluster, is usually set as the origin of the coordinate system and all angular positions of multiple images are determined with respect to it.

One of the simplest examples for case (2) in Section~\ref{sec:breaking_degeneracies} to break the degeneracy in Equation~\eqref{eq:H0_degeneracy} by a local measurement independent of $H_0$ employs a singular isothermal elliptical mass density model (SIE), as introduced in \cite{bib:Kormann}. 
In this case, $\psi(\boldsymbol{x},f)$ with the axis ratio $f=b/a$ of the semi-minor axis $b$ to the semi-major axis $a$ of the elliptical critical curve. 
Being able to measure $a$ and $b$ for a given multiple-image configuration, e.g.\ from an ``Einstein ellipse" around the critical curve, we can determine $\Delta \phi(\boldsymbol{x}_1,\boldsymbol{x}_2,f)$ from $\psi(\boldsymbol{x},f)$ as the left-hand side of Equation~\eqref{eq:invariance_transformation}. 
Inserting $\Delta \phi(\boldsymbol{x}_1,\boldsymbol{x}_2,f)$ into Equation~\eqref{eq:H0_degeneracy} subsequently yields $H_0$. 

A simple example for the case (3) in Section~\ref{sec:breaking_degeneracies} to break the degeneracy in Equation~\eqref{eq:H0_degeneracy} by adding a second equation that depends in another way on $\Delta \phi(\boldsymbol{x}_1,\boldsymbol{x}_2,\boldsymbol{p})$ and $H_0$ is the mass density model of a circularly symmetric lens. 
Observing an Einstein ring with the angular Einstein radius $r_\mathrm{E}$ around a deflecting circular mass $M$ that is enclosed within the radius $r_\mathrm{E}$, we have
\begin{align}
\tau &= \dfrac{1+z_\mathrm{l}}{c} \dfrac{4 G M}{c^2} \left( \dfrac12 \left(\boldsymbol{x}_1 - \boldsymbol{x}_2 \right)^\top \left(\dfrac{\boldsymbol{x}_1}{\boldsymbol{x}_1^2} + \dfrac{\boldsymbol{x}_2}{\boldsymbol{x}_2^2} \right) - \ln \left( \dfrac{\boldsymbol{x}_1}{\boldsymbol{x}_2} \right) \right) \\
r_\mathrm{E} &= \sqrt{\dfrac{4  G M}{c^2} \dfrac{D_\mathrm{ls}}{D_\mathrm{s} D_\mathrm{l}}} \;, \label{eq:Einstein_ring}
\end{align}
such that combining these equations yields
\begin{align}
H_0 = \dfrac{\tilde{\Gamma}}{\tau} r_\mathrm{E}^2  \left( \dfrac12 \left(\boldsymbol{x}_1 - \boldsymbol{x}_2 \right)^\top \left(\dfrac{\boldsymbol{x}_1}{\boldsymbol{x}_1^2} + \dfrac{\boldsymbol{x}_2}{\boldsymbol{x}_2^2} \right) - \ln \left( \dfrac{\boldsymbol{x}_1}{\boldsymbol{x}_2} \right) \right) \;,
\end{align}
which has already been proposed by \cite{bib:Refsdal}. \com{A similar equation to Equation~\eqref{eq:Einstein_ring} was set up by \cite{bib:Witt} to constrain $H_0$ for the class of generalised isothermal lens models. They also note that the source can be eliminated from the description for a special sub-class of models.}
However, \com{apart from \cite{bib:Witt}}, compared to previous derivations to determine $H_0$, we are not aware of any \com{further approaches} that are independent of the source, as these equations are.
\com{Possibly, this is the case because previous approaches have considered the time delay as a function of a position in the source plane and a gravitational deflection potential on a cosmological background such that the multiple images are the stationary points of the time delay, i.e.\@ the positions that fulfil Fermat's principle of extremal travel time.
This rather model-based viewpoint focuses on the generation of multiple images from any point $\boldsymbol{y}$ in the source plane, while our data-driven approach aims at a description of already observed multiple images which fulfil the lens equation.}

For galaxy-scale gravitational lens modelling, the elimination of the source is of minor \com{advantage for those} lens model methods \com{that} are based on maximising the overlap between the back-projected multiple images into the source plane. 
Even when eliminating the source from Equation~\eqref{eq:time_delay41}, it has to be reconstructed to obtain a self-consistent lens model. 
Furthermore, on galaxy scale, time delay differences are on the order of days to months.
Hence, they are often observable between all multiple images of a background source.
Contrary to that, the elimination of the source from Equation~\eqref{eq:time_delay41} for cluster-scale lenses can be advantageous, as time delay differences are on the order of years and therefore may not be available between all multiple images of a background source. (This can especially happen for transient events like supernovae or fast radio bursts.)
Equation~\eqref{eq:time_delay41} has to be coupled to the back-projection of the multiple images, which is usually realised by replacing $\boldsymbol{\alpha}(\boldsymbol{x})$ by $(\boldsymbol{x} - \boldsymbol{y})$ in Equation~\eqref{eq:time_delay3}. 
As already noted in \cite{bib:Wagner7}, inserting the source position into Equation~\eqref{eq:time_delay3}, several choices arise with respect to the implementation of the coupling. 
For instance, we could decide to reconstruct the source to be inserted into  Equation~\eqref{eq:time_delay3} by back-projecting all multiple images and maximising their overlap.
Alternatively, as Equation~\eqref{eq:time_delay3} only connects multiple images with time delay differences, we could restrict the source reconstruction for Equation~\eqref{eq:time_delay3} to these images without taking into account the additional images without observed time delay differences.
Employing Equation~\eqref{eq:time_delay41}, these implementational ambiguities are avoided. 
Whether this also leads to an increase in precision and accuracy of the lens model compared to using the formulation including the source is subject to current investigations.

In general, if Equation~\eqref{eq:H0_degeneracy} is supposed to break the $H_0$-$\Delta \phi$-degeneracy, we require $n_\mathrm{p}$ additional equations with independent observables.
Altogether, the $n_\mathrm{p} + 1$ equations thus determine the $n_\mathrm{p}$ parameters of $\psi(\boldsymbol{x},\boldsymbol{p})$ and $H_0$.
Employing a \emph{global} lens model $\psi(\boldsymbol{x},\boldsymbol{p})$ breaks the degeneracy in Equation~\eqref{eq:H0_degeneracy} because the global parametrisation by $\boldsymbol{p}$ imposes the missing constraints to connect Equation~\eqref{eq:H0_degeneracy} with further constraining equations like Equation~\eqref{eq:Einstein_ring}.  
Yet, assuming a global lens model may also introduce biases. 
As soon as the model has less parameters than there are degrees of freedom in the true underlying mass density distribution, biases arise if these systematic oversimplifications are larger than the confidence intervals given by the measurement precision.
Such biases are, for instance, investigated in \cite{bib:Xu_H0} and \cite{bib:Sonnenfeld} for the determination of $H_0$ from multiple images in simulated galaxy-scale lenses of varying degree of complexity and model complexity for the reconstruction of $\Delta \phi$.

On the other hand, it is also possible to introduce much more parameters than actually necessary to capture the complexity of the deflection potential up to our measurement precision. 
These parameters can be degenerate among each other, as, for instance, investigated in \cite{bib:Suyu3} and \cite{bib:Suyu4} when determining $H_0$ from galaxy-scale lenses of the H0LiCOW sample.
The quality of fit of such models to the observations shows how well the complexity of the data is captured employing the optimum model parameter set. 

Another way to globally reconstruct the mass density distribution is free-form modelling, describing the mass density distribution (or the deflection potential) as a superposition of a number of basis functions, see e.g. \cite{bib:Williams3}, \cite{bib:Liesenborgs1}, \cite{bib:Merten}.
Contrary to specific density profiles with usually few parameters, the number of adjustable parameters can be much higher in free-form modelling, if the model comprises a lot of basis functions.
Allowing for local fine-tuning, e.g. by adding masses in regions without multiple images, an entire class of globally differing lens models that fulfil Equation~\eqref{eq:invariance_transformation} can be generated in free-form approaches, see \cite{bib:Liesenborgs3}. 
The quality of fit of such parameter-free models to the sparse amount of observations is given by the size of the constraint set of all possible solutions. 
The more constraining the data, the smaller the set and the tighter fit. 
This relation is investigated in \cite{bib:Williams1} for the determination of $H_0$ from the multiple images of supernova Refsdal in the galaxy-cluster-scale lens MACS~1149. 
Constraints from additional assumptions, such as symmetry or constraints on the radial slope of the density profile, can decrease the set of equally suited lens reconstructions as well, until $H_0$ is determined to the desired precision. 

Hence, both parametric and free-form lens models can be employed to break the $H_0$-$\Delta \phi$-degeneracy.
Due to their complementary modelling approaches, it seems most appropriate to determine $H_0$ with both methods using the same observations and adding the same additional assumptions in the free-form approach and the parametric model.
The value for $H_0$ should be consistently obtained. Yet, most importantly, the confidence interval for $H_0$ as obtained by the parametric model can be systematically disentangled into the constraining power of the data and the individual additional assumptions by means of the free-form approach.



\subsection{Breaking degeneracies by non-lensing observables (velocity dispersions)}
\label{sec:observation_breaking}

A second way to determine a global reconstruction for $\psi(\boldsymbol{x})$, for all $\boldsymbol{x} \in \mathcal{X}$, to infer $\Delta \phi(\boldsymbol{x}_1, \boldsymbol{x}_2)$ is employing measurements of the velocity dispersions along the line of sight of the luminous part of the deflecting mass distribution.
If the deflecting object is a galaxy, the velocity dispersions of the stars are used. 
Since most strong lensing galaxies are too far from us to resolve the individual motions and dispersions of the stars within the lens plane, only the velocity and the velocity dispersion along the line of sight can be determined from spectra. 
Depending on the resolution of the spectrograph, we can obtain a single spectrum for the lensing galaxy.
From this, a luminosity weighted velocity dispersion along the line of sight can be derived, as, for instance, employed in \cite{bib:Koopmans}, \cite{bib:Suyu2}, \cite{bib:Wong1}, or \cite{bib:Birrer2}.
These measurements yield a luminosity weighted velocity dispersion along the line of sight integrated in the limits of an aperture, while models may contain central velocity dispersions. 
As argued in \cite{bib:Koopmans1}, it is more convenient to work with the velocity dispersion values obtained from the observables and reformulate the models, instead of converting the data into central values according to \cite{bib:Treu}. 
With modern integrated field spectrographs, it has become possible to observe luminosity weighted spectra on angular scales smaller than a galaxy, as, for instance, employed in \cite{bib:Barnabe}, \cite{bib:Czoske}, or \cite{bib:Sonnenfeld1}.

For galaxy clusters acting as deflecting objects, the luminosity weighted velocity dispersions along the line of sight for the cluster member galaxies are observed.

The stars in galaxies and the galaxies in clusters can be characterised as collisionless ensembles of test particles. 
As an usual approximation, all test particles are assumed to be identical and to be moving in a three-dimensional gravitational potential $\psi_{3d}(\boldsymbol{l})$, $\boldsymbol{l} \in \mathbb{R}^3$.
The latter is caused by the total (luminous and dark) local matter density distribution and as such independent of the background density and thus $H_0$.
Hence, employing the velocity dispersions to determine $\psi(\boldsymbol{x})$ is an example for case (2) of Section~\ref{sec:breaking_degeneracies}.

Setting up the Jeans equations to describe the three-dimensional motion of the test particles in the external field of $\psi_{3d}(\boldsymbol{l})$,
\begin{align}
\partial_t \boldsymbol{v}(\boldsymbol{l},t)+ (\boldsymbol{v}(\boldsymbol{l},t) \nabla_{\boldsymbol{l}}) \boldsymbol{v}(\boldsymbol{l},t) + \dfrac{\nabla_{\boldsymbol{l}} \left( S n(\boldsymbol{l},t) \right)}{n(\boldsymbol{l},t)} = - \nabla_{\boldsymbol{l}} \psi_{3d}(\boldsymbol{l}) 
\label{eq:Jeans}
\end{align}
relates spatial derivatives of $\psi_{3d}(\boldsymbol{l})$ to the mean velocities of the test particles and the divergence of the tensor $S$ of the luminosity weighted velocity dispersions.
$\boldsymbol{v}(\boldsymbol{l},t)$ denotes the vector of mean velocities in the three spatial directions at $\boldsymbol{l}$ at time $t$, $S \in \mathbb{R}^{3\times 3}$ contains the velocity dispersions $(\sigma^2)_{ij}$, $i,j=1,2,3$, at $\boldsymbol{l}$ at time $t$, and $n(\boldsymbol{l},t)$ denotes the mean number density of test particles at $\boldsymbol{l}$ at time $t$.

Since only the measurement of the luminosity weighted velocity dispersion along the line of sight at given $\boldsymbol{l}$ at time $t$ is available, Equation~\eqref{eq:Jeans} cannot be solved without employing further assumptions.
Usually, we assume the system to be in a steady state equilibrium, so that the time derivative vanishes.
In addition, the velocity components in the different directions are assumed to be statistically independent and that there is no mean velocity in any direction, $\boldsymbol{v}(\boldsymbol{l}) = 0$.
It follows, that the only non-vanishing components of the velocity dispersion tensor are on the diagonal. 
Expressed in spherical coordinates $(r,\varphi, \vartheta)$, the non-vanishing velocity dispersion components are assumed to be related by
\begin{align}
\sigma_\varphi^2 = \sigma_\vartheta^2 =  (1-\beta(r)) \sigma_r^2 \;,
\end{align}
introducing the anisotropy parameter $\beta(r)$ for the distribution of velocities.
(If the distribution of velocities is isotropic, $\beta(r) = 0$.)
Under these requirements, Equation~\eqref{eq:Jeans} simplifies to
\begin{align}
\partial_r \left( n(r) \sigma_r^2 \right) + \dfrac{2 \beta(r)}{r} n(r) \sigma_r^2 = - n(r) \partial_r \psi_{3d}(r) \;.
\label{eq:Jeansr}
\end{align}
Hence, connecting the mean number density of test particles, $n(r)$, to the luminosity profile of the lens, Equation~\eqref{eq:Jeansr} constrains the radial slope of $\psi_{3d}(r)$, if $\beta(r)$ is also known or set to zero for an isotropic distribution of velocities. 
By means of the Poisson equation, $\partial_r \psi_{3d}(r) = G M_{3d}(r)/r^2$ can be inserted into the left-hand side of Equation~\eqref{eq:Jeansr} to relate the right-hand side to the total mass of the system. 
If $\beta(r) \ne 0$, Equation~\eqref{eq:Jeansr} is subject to the so-called mass-anisotropy degeneracy, relating the unknown anisotropy parameter $\beta(r)$ and the mass $M_{3d}(r)$ with each other that cannot be both determined from Equation~\eqref{eq:Jeansr}.

Subsequently projecting $\psi_{3d}(r)$ or $M_{3d}(r)$ along the line of sight yields the two-dimensional $\psi(\boldsymbol{x})$ or $M(\boldsymbol{x})$ employed in  Equation~\eqref{eq:H0_degeneracy}.
Yet, as the angular positions at which the velocity dispersions are measured cannot coincide with the angular positions of the multiple images, Equations~\eqref{eq:H0_degeneracy} and \eqref{eq:Jeansr} cannot be directly coupled to each other. 
The introduction of a model for $\psi_{3d}(r)$ can help. 

Employing a certain model in Equation~\eqref{eq:Jeansr} and its projected version in Equation~\eqref{eq:H0_degeneracy}, we arrive at a consistent coupling. 
This was realised e.g. in  \cite{bib:Barnabe}. 
They showed for their model-based coupling approach that the degeneracy in Equation~\eqref{eq:H0_degeneracy} can be broken by adding the constraints from the velocity dispersions when using the same gravitational potential for both equations.

Measurements of velocity dispersions, with relative imprecisions on the order of 5-10\%, are still comparably imprecise compared to characteristics of multiple images\footnote{Except for magnification ratios of multiple images, which are rarely used for this reason \textbf{due to contaminations from microlensing or dust extinction}.}. 
Therefore, using velocity dispersions to break the degeneracy in Equation~\eqref{eq:H0_degeneracy} comes at the cost of introducing high imprecisions apart from an additional set of assumptions and the need for a model to couple both observations. 

For most cases, the imprecisions in the velocity dispersions are large enough that gravitational lensing and test particle dynamics can be separated and different potentials can be used. 
Usually, elliptical lens models are combined with spherical models to be reconstructed from the velocity dispersions, such that their radial profiles are the same. 
This ansatz is, for instance, pursued in \cite{bib:Koopmans}, \cite{bib:Suyu2}, \cite{bib:Sonnenfeld1}, \cite{bib:Wong1}, or \cite{bib:Birrer2}. 
\cite{bib:Sonnenfeld1} demonstrated that current measurement precisions of velocity dispersions do not allow to constrain the ellipticity of the gravitational potential for galaxies yet.
Furthermore, \cite{bib:Birrer3} showed that the specific model, i.e. the choice of $\beta(r)$, and the prior of the stellar dynamics part contributes most to the overall error budget when inferring $H_0$ for the galaxy-scale lens RXJ1131-1231.

Since time delay measurements of multiple images in galaxy clusters are rare, only one cluster-scale lens has been used to determine $H_0$ so far, \cite{bib:Vega}, \cite{bib:Grillo}, \cite{bib:Williams1}. 
The degeneracy in Equation~\eqref{eq:H0_degeneracy} is broken by lens modelling employing a multitude of multiple-image systems from sources at difference redshifts, no additional observables are required. 
It can be doubted that this ansatz breaks the mass sheet degeneracy, as noted in \cite{bib:Liesenborgs1}. 
But for a globally parametrised lens model, it is possible to break the mass sheet degeneracy assuming a high degree of smoothness of the mass density distribution.
The latter is essential to prevent the algorithm from overfitting. 
Alternatively, the three-dimensional gravitational potential can also be reconstructed from cluster member velocity dispersion measurements, as shown in \cite{bib:Stock}, in case a coupling of dynamics and lensing on cluster scale may become necessary in the future. 

 
\begin{figure*}
\centering
\includegraphics[width=0.45\textwidth]{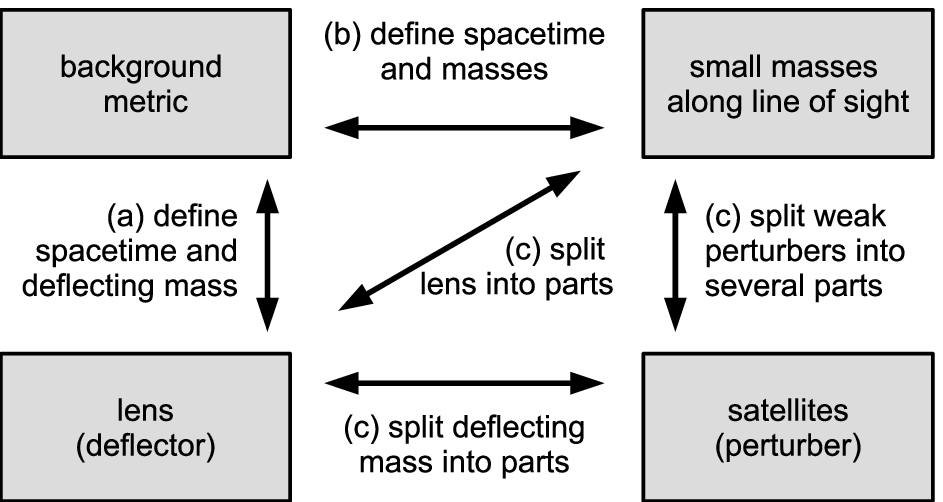}
\hfill
\includegraphics[width=0.45\textwidth]{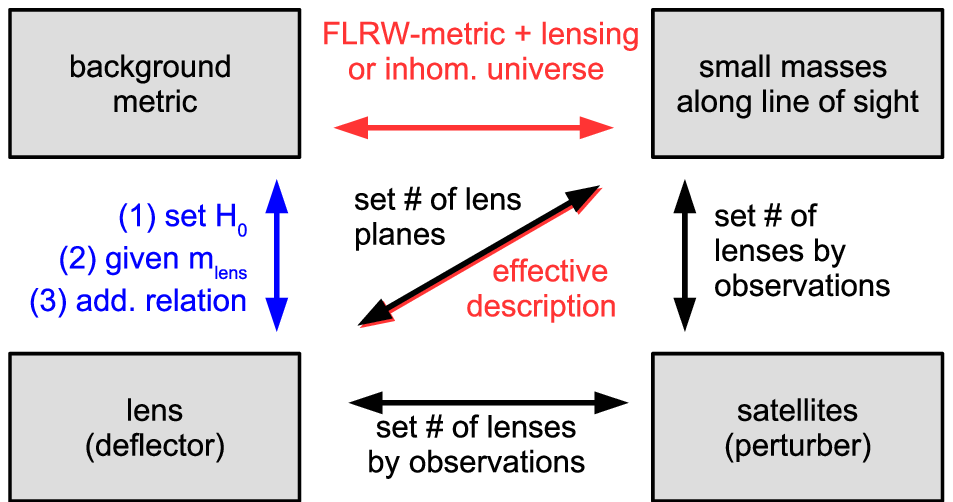}
\caption{Left: Summary of all degeneracies arising in the standard gravitational lensing formalism as detailed in Section~\ref{sec:principles}; Right: Summary of necessary conditions to break the degeneracies discussed in Section~\ref{sec:breaking_degeneracies} (blue), choices to be set to break degeneracies (red), and observations to be included to break degeneracies (black) as discussed in Section~\ref{sec:derivation}.}
\label{fig:synopsis}
\end{figure*}

\subsection{Summary of degeneracy breaking}
\label{sec:summary} 
Summarising the results, we observe that the usage of a globally parametrised lens model can break the degeneracy in Equation~\eqref{eq:H0_degeneracy}. 
To determine the model parameters, we have to employ further observables of the multiple image configuration(s), apart from the time delay. 
The most information can be gained from non-local observables as Einstein rings or giant arcs close to the critical curves. 
These extended observations put stronger constraints on the symmetry of the mass density distribution than multiple images that are farther away from the critical curve and sparsely distributed. 
Furthermore, all lens models coincide in the position of the critical curves, so that model-based biases should decrease with increasing proximity of the observables to the critical curves. 
Yet, since we do not know the complexity of the true mass density, introducing a lens model still comes at the cost of potential biases if the model is oversimplified. 
On the other hand, introducing a lot of lens model parameters may cause degeneracies among themselves, which may lead to overfitting. 

Adding an observation of a luminosity weighted velocity dispersion along the line of sight, it is possible to obtain an estimate of the deflection potential independent of the background density. 
Yet, coupling gravitational lensing to the dynamics of the luminous matter in the overall gravitational potential of the total deflecting mass comes at the cost of employing comparably imprecise velocity dispersion measurements. 
It also requires a lot of additional assumptions and a model to extrapolate the reconstructed and projected deflection potential from dynamics to the position of the multiple images.

While the ensemble of currently available observational cases yields $H_0$ to percent precision, \cite{bib:Birrer2}, further investigations into improving the precision of the velocity dispersion measurements and constraining $\beta(r)$ are highly desired to further increase the precision of $H_0$ for the individual lenses. Complementarily, improving the estimates of the differences in the Fermat potential derived from lens models can also increase the precision of $H_0$.

\section{Conclusion}
\label{sec:conclusion}

We investigated the physical causes of the degeneracies that enter the gravitational lensing formalism and thus occur for gravitational lensing on galaxy and on galaxy cluster scale. 
We found that they appear for two reasons, as summarised in Figure~\ref{fig:synopsis} (left) and detailed in Section~\ref{sec:summary_of_degeneracies}.
The first one is the freedom to partition the total matter density distribution along a line of sight into an overall background, small-scale perturbations, a main deflector, and potential satellites. 
The second one is the fact that, except for the time delay difference, observables obtained from multiple images are angular quantities within the lens plane.
This implies that only ratios and reduced lens characteristics can be determined from them without imposing further assumptions.

As usual, we adopted a FLRW metric to determine the distances to the lens and the source. 
Any (small-scale) matter densities were treated as (weak) lenses, although they could be equivalently absorbed into a more complicated metric with modified distance-redshift relations.
We furthermore restricted our analyses to the effective single-lens plane formalism.
This is sufficient to characterise most known cases because they do not show rotations between multiple images that require a multi-lens-plane description.
Moreover, detailed observations of the lens environment and its mass density distribution along the line of sight are often not available.

Given these prerequisites, we found that $H_0$ can be related to the overall energy content of the universe today, such that it can be interpreted as the background density on top of which gravitational lensing can be defined, see Equation~\eqref{eq:H0_density}.
Consequently, the degeneracy between the $\Delta \phi$ and $H_0$ in the time delay equation, as arising in Equation~\eqref{eq:H0_degeneracy}, can be interpreted as the freedom to consistently rescale the background and the deflecting mass on top of it without changing the observables of the multiple images.
We showed that, from a mathematical and a physical point of view, no further degeneracies arise in this equation.

Reformulating the time delay equation as done in Equation~\eqref{eq:time_delay41}, it only depends on observables and the lens.
Being independent of the source simplifies the treatment of degeneracies, but also allows for a more efficient algorithmic implementation for mass density reconstructions of the lens and the inference of $H_0$, as we will show in a future work.

To determine $H_0$ from Equation~\eqref{eq:H0_degeneracy}, choosing highly symmetric configurations of multiple images close to critical curves is important because it allows for a simple lens model and being close to the model-invariant critical curve reduces model biases. 
Since time delay differences decrease with increasing proximity to the critical curve, the precision to which they can be determined is an important factor. 
With the recently discovered fast radio bursts, an additional time-varying source has become available, which has duration times as short as milliseconds. 
 
Breaking the degeneracy between $\Delta \phi$ and $H_0$ by lens models or by including velocity dispersions along the line of sight, we arrived at the results summarised in Section~\ref{sec:summary}. Figure~\ref{fig:synopsis} (right) gives a diagrammatic summary how to break this and the other degeneracies listed in Figure~\ref{fig:synopsis} (left).
Both approaches introduce additional assumptions and observables of an extended region and break the degeneracy on a global level.
As a consequence, the added models have to be chosen to avoid oversimplification biases or overfitting, and to minimise the number and degeneracies of model parameters.
It would suffice to break the degeneracy locally, as the time delay equation is confined to the positions of the multiple images. 

The currently best approach, employing lens models and velocity dispersions, \cite{bib:Suyu}, shows that percent precision in $H_0$ can be achieved by compiling several cases on galaxy scale. 
Potentially, a further increase in the precision of the $H_0$ obtained for the individual cases can be achieved by employing the source-independent time delay equation as formulated in Equation~\eqref{eq:time_delay41}. 
Marginalising over the density parameters $\Omega_i$ could also be replaced by employing the data-based distances as developed in \cite{bib:Wagner5} because all lenses and sources except for the source of HE 1104-1805 ($z_\mathrm{s} = 2.316$) lie in the range of the Pantheon data set ($z < 2.3$).

The major contribution to the confidence bounds on $H_0$ comes from the uncertainties of the velocity dispersions and further constraints from additional multiple images on lens models on galaxy scale are rare. 
Therefore, we also recommend the inference of $H_0$ from multiple images in galaxy clusters. 
Available cases are still rare, yet, multiply-imaged quasars or supernovae behind clusters may be detected more frequently with future surveys.
As investigated in \cite{bib:Wagner7}, fast radio bursts could be equally likely found as multiple images in a galaxy cluster.
Advantages in favour of employing cluster-scale lenses are the higher relative precision of the time delay measurements and the numerous constraints on the lens model coming from a multitude of multiple-image systems. 
As state-of-the-art cluster reconstruction algorithms were shown to become accurate and precise to percent level in reconstructing the mass enclosed within the critical curves, see \cite{bib:Meneghetti}, a cluster lens model may lead to a higher precision and accuracy for $H_0$ than achievable by including the currently best velocity dispersion measurements on galaxy scale. 
Furthermore, using gravitational lensing alone is a more direct probe of the dark matter part of the deflecting mass density, as no assumptions about the galaxy dynamics within the cluster are necessary. 
\com{Future simulations similar to those of \cite{bib:Meneghetti}, yet to be performed, will show, whether the reconstruction of the Fermat potential achieves an accuracy and precision that surpasses those of galaxy-scale lenses. A simulation-based investigation of the reconstruction accuracy and precision for the latter is currently being performed by \cite{bib:Ding}.}

\section*{Acknowledgements}

I would like to thank Matthias Bartelmann, Simon Birrer, Marc Gorenstein, Bettina Heinlein, Michael F. Herbst, Bruno Leibundgut, Jori Liesenborgs, Sven Meyer, Henrik Nersisyan, Cristian Eduard Rusu, Dominique Sluse, Sebastian Stapelberg, Nicolas Tessore, R\"udiger Vaas, Olivier Wertz, Liliya Williams, and the anonymous referee for helpful discussions and comments. I gratefully acknowledge the support by the Deutsche Forschungsgemeinschaft (DFG) WA3547/1-1 and WA3547/1-3. 




\bibliographystyle{mnras}
\bibliography{mnras} 



%

\bsp	
\label{lastpage}
\end{document}